\documentclass[draft]{aipproc}

\layoutstyle{6x9}

\def\be{\begin{equation}}
\def\ee{\end{equation}}
\def\bea{\begin{eqnarray}}
\def\eea{\end{eqnarray}}

\begin{document}

\title{String bits and the Myers effect}

\author{Pedro J. Silva}{
  address={Physics Department, Syracuse University, Syracuse, New York 13244}
}

\begin{abstract}
Based on the non-abelian effective action for D1-branes, a new
action for matrix string theory in non-trivial backgrounds is
proposed. Once the background fields are included, new interactions
bring the possibility of non-commutative solutions i.e. The Myers effect
for ``string bits'' .
\end{abstract}

\maketitle


\section{Matrix strings}

Matrix string theory \cite{dvv,bs,cj1} is one of the most interesting outcomes
of the different dualities in M-theory. Perhaps a simple way to define it is
by looking at Matrix theory \cite{bfss} with an extra compactified dimension.
For example, following Dijkgraaf et. al. \cite{dvv}, if we
consider M-theory on a torus of radii $A$ and $B$, by first
reducing  on $A$  and then making an infinite boost on $B$ we get
type IIA string theory on the discrete light cone (DLC) with
D0-brane particles. If, on the other hand, we reduce on $B$ first,
boost and then consider t-duality on $A$, we get (1+1) super
Yang-Mills (SYM) theory with fundamental string charge on the
world-volume i.e. the low energy theory of D1-branes. One finds therefore,
 that Matrix string theory is a non-perturbative
definition of string theory built in terms of a two dimensional
SYM theory and a collection of scalar fields in the adjoint
representation of the gauge group (see the original papers for an
extended discussion of this derivation). Although we have
discussed only type IIA string theory, there are other
constructions similar to the one sketched before, where the other
four superstring theories are written in terms of two dimensional
SYM theory\footnote{Actually for the cases of type IIB and type I
string theory there is an additional construction in terms of a
three dimensional theory \cite{cj1}}.

The Matrix string theory conjecture was originally formulated on
flat backgrounds. Lately, using some techniques developed by
Taylor and Raamsdonk \cite{tvr1} a generalization for closed
strings on non-trivial weak backgrounds has appeared \cite{sch}.
This talk is based on the paper \cite{sil}, a further
generalization of the original Matrix string theory to non-trivial
weak backgrounds based on the non-abelian D1-brane action proposed
by Myers.

In \cite{sch,sil}, the possibility of non-abelian configurations
of fundamental strings was pointed out. In particular, the
appearance of a Myers-like effect was computed explicitly (By now
the Myers effect \cite{mye1} is a well known phenomenon where $N$
D-branes adopt a non-abelian configuration that can be understood
as a higher dimensional abelian D-brane). These configurations
come about as the result of new interaction terms that appear in
the non-abelian effective actions.

The appearance of strings describing D-branes is not new. There
are computations of Dp-branes collapsing into fundamental strings
\cite{mal} and fundamental strings blowing up into Dp-branes
\cite{rem}, always in terms of the abelian Born-Infeld actions of
the corresponding D-branes. What is new in the matrix string
formulation is that we have a formalism in which a two-dimensional
action naturally includes matrix degrees of freedom representing
the ``string bits''\footnote{The idea is that the string can be
seen as a chain of partonic degrees of freedom \cite{tho1}}, which
also incorporate the description of higher dimensional objects of
M-theory using non-commutative configurations.

One of the important properties of this new theoretical framework
lies in the similarity of the mathematical language used to
describe the fundamental objects of M-theory, bringing for first
time the possibility of describing strings and D-branes in a
unified framework, a ``democracy of p-branes'' \cite{twn1}.

\section{Matrix string and Non-abelian D1-branes}

In the previous section we talked about a theoretical framework that
describes fundamental strings in terms of matrix degrees of freedom.
For example, in type IIA this action is a two dimensional
supersymmetric gauge theory that contains DLC string theory and
has extra degrees of freedom representing non-perturbative objects
of string theory. Also, it is a second quantized theory as it is
built from many strings.

We know that by means of different dualities the five superstring
theories are described in the neighborhood of a 1+1 dimensional
orbifold conformal field theory. In this language the strings are
free in the conformal field theory limit, representing DLC string
theory. The interactions between the strings are turned on by
operators describing the splitting and joining of fundamental
strings. These operators deform the theory away from the conformal
fixed point.

To further clarify these ideas, let us follow a sketch of
the derivation for the case of type IIA string theory.
Consider type IIA strings in the DLC frame with
string mass $m_A$, string coupling $g_A$ and a null compact
direction of radius $R_A$ (where we identify the null coordinate
as $x^- \approx x^- + R_A$).

Using the relation between the null compactification and a space-like
compactification a la Seiberg-Sen \cite{ss}, we get type IIA
string theory on a space-like circle of radius $R$ in the sector
with momentum $N$, string mass $m$ and  string coupling $g$. The
relation between these two heterotic string theories is given by
\be
m^2R=m_A^2R_A\;\;,\;\;g=g_A\;\;,\;\;R\rightarrow 0.
\ee
Next, we perform a t-duality transformation on $R$, so that the new constants of the
string theory $(m',g',R')$ are given by
\be
m'=m\;\;,\;\;g'={g\over mR}\;\;,\;\;R'={1\over m^2R}.
\ee
Finally, we perform a s-duality transformation to obtain type IIB string theory with $N$
D1-strings and constants $(m_b,g_b,R_b)$ given by the following
expressions,
\be
m_b={m'\over g'^{1/2}}\;\;,\;\;g_b={1\over
g'}\;\;,\;\;R_b=R'.
\ee
In terms of the initial type IIA
theory and $R$ we get
\bea
&&m_b=\left[{m_A^{6}R_A^{3}\over g_A^{2}R}\right]^{1/4} \rightarrow \infty,\nonumber\\
&&g_b=\left[{m_A^2R_AR\over g_A^2}\right]^{1/2} \rightarrow 0,\nonumber\\
&&R_b={1\over m_A^2R}.
\eea
Therefore, we get the low energy
theory of N D1-branes at weak coupling, where the gauge coupling
constant $g_{YM}$ is given by,
\be
g_{YM}\propto m_A^2R_A/g_A.
\ee
This is the 1+1 dimensional SYM theory with eight scalars in the
adjoint representation of the gauge group. This effective action
is obtained by the dimensional reduction of $N=1$ supersymmetry
Yang-Mills theory in ten dimensions down to two dimensions.

To define the type IIB case, a possible route to take is to start with
type IIB strings in the DLC frame, then perform a t-duality transformation on
the null circle taking us to type IIA in the DLC. This is similar to the
previous situation with the difference that winding modes are exchanged
for momentum modes. The relation between the corresponding meaningful constants is,
\bea
&&m_b=\left[{m_B^2\over g_B^2R_BR}\right]^{1/4}\rightarrow \infty,\nonumber\\
&&g_b=\left[{m_B^2R_BR\over g_B^2}\right]^{1/2}\rightarrow 0,\nonumber\\
&&R_b=R_B, \label{eq:b1}
\eea
where $(m_B,g_B,R_B)$ are the
initial type IIB string parameters and $(m_b,g_b,R_b)$ are the
final (also type IIB) string theory parameters. Again, we get a
low energy weakly coupled string theory with N D1-branes. The gauge
coupling constant $g_{YM}$ is given by
\be
g_{YM}=m_B/g_B. \label{eq:b2}
\ee
The heterotic case is similar but some care has to be taken with the
inclusion of Wilson lines \cite{kro}. On the other hand, type I theory is more subtle
and is related to the low energy limit of type IA theory in the presence of D8-branes and
D0-branes plus winding modes on the orbifold. Therefore it is a quantum mechanics system
but with an infinite tower of winding modes.

In order to obtain the relevant action for one of the five matrix
string theories, we start with the world-volume gauge theory of N
D1-branes, and then go back along the chain of dualities until we
reach the desired DLC string theory. For example, consider first
an s-duality transformation on the D1-brane effective action, then
a t-duality transformation and finally the boost relations of
Seiberg-Sen. As a result we get type IIA matrix string theory.
This can be written as
\be
L^{IIA}_F\equiv B\circ T\circ S
\;[L_{D1}]. \label{eq:3}
\ee
Other matrix string theories
Lagrangians can be obtained by similar procedures. For example,
$L^{IIB}_F\equiv T\circ B\circ T\circ S \;[L_{D1}]$.

As we mentioned in the introduction, there are generalizations of
the matrix string action which include weak backgrounds. This time
the calculations are based on the relation between matrix
string and the matrix theory proposal. In particular, previous
works of Taylor and Van Raamsdonk \cite{tvr1} are used to support
these results. One of the positive outcomes of the above work is a
proposal for the transformation of the D1-brane world-volume
fields under s-duality. Thus, based on these different proposals
we are able to actually construct the matrix actions using maps
like the one in equation (\ref{eq:3}).

It is important to note that recently Myers wrote a non-abelian
action of N Dp-branes in general backgrounds \cite{mye1} which is
fully covariant under t-duality. This action incorporates (in the
limit of weak backgrounds), all the couplings derived previously by
Taylor et. al. and also introduces some new ones. If we believe
this effective action for the D1-branes, we are forced to
conjecture that:

{\it Matrix string theory is defined by Myers D1-brane world-volume action
plus the web of dualities needed}.

Note that since the non-abelian D1-brane action proposed by Myers
does not capture the full physics of the infrared limit, we can
only trust its expansion up to sixth-order in the field strength
\cite{bai}, and this problems is inherited by the above conjecture
for the matrix theory action. Another technical problem comes from
the chain of dualities, since it makes it difficult to give an
explicit closed form for the final Lagrangian. In particular, the
t-duality map mixes RR fields and NS fields. Nevertheless, we only
have to use the Buscher rules \cite{bus} on the supergravity
background fields as t-duality (once we have s-dualized), leaves
the world-volume fields invariant. At last the action of Myers only
tells us about the bosonic degrees of freedom, therefore the
fermionic counterpart has to be calculated using supersymmetry.

For example, let us consider the type IIA case. Following equation
(\ref{eq:3}), the action for the Matrix string is given
in two parts, the first corresponding to the original Born-Infeld
term of the D1-brane action,
\bea
S^{1}_{F1}={1\over\lambda}\int d\xi^2 Str
\left\{
\sqrt{-det(P[\widetilde{E}+\widetilde{E}(\widetilde{Q}^{-1}-\delta)\widetilde{E}]+
\lambda
e^{\widetilde{\phi}}\widetilde{g}F)det(\widetilde{Q})}\right\}
\label{eq:4}
\eea
where
\bea
&&\widetilde{E}_{AB}=\widetilde{G}_{AB}-e^{\widetilde{\phi}} \widetilde{C}^{(2)}_{AB},\nonumber \\
&&\widetilde{Q}^{\;i}_{\;j}=\delta^i_{\;j}+i\lambda[\Phi^i,\Phi^k]\widetilde{E}_{kj}
(\widetilde{g}e^{\widetilde{\phi}})^{-1},
\eea
and the tilde represents the t-dual transformation of the background
fields. For example the form of $\widetilde{C}_{AB}$ is
\bea
&\widetilde{C}_{AB}=\left( \begin{array}{cc} C_{\alpha\beta
y}+2C_{[\alpha }B_{\beta]y}-2C_yB_{y[\alpha}G_{\beta]y} &
\;\;\;C_\alpha-C_yG_{\alpha y}/G_{yy}\\
-C_\beta+C_yG_{\beta y}/G_{yy}& 0\end{array}\right)& ,
\eea
where the space-time index $A$ has been divided into the t-dualized
direction $y$ and the other directions $\alpha$.

The second part, corresponding to the original Chern-Simons term
of the D1-brane action is
\bea
S^{2}_{F1}={1\over
\lambda}\int d\xi^2 STr\left\{ P\left[
e^{i\widetilde{g}^{-1}\lambda\,i_\Phi
i_\Phi}[(-\widetilde{B}+\widetilde{C}^{(4)})
e^{-\widetilde{C}^{(2)}}] \right] \, e^{\lambda
\widetilde{g}F}\right\}. \label{eq:5}
\eea
This action contains the action of the matrix string theory of Dijkgraaf
et. al. \cite{dvv}, since by construction in trivial backgrounds the
D1-brane action of Myers gives the 1+1 SYM theory corresponding
the dimensional reduction on N=1 SYM in ten dimensions down to two
dimensions. Hence, by taking all of the background fields to be
trivial, we recover the standard form of type IIA matrix string theory,
\be
S^{1}_{F1}=\lambda\int d\xi^2 Tr
\left\{ {\partial\Phi^2\over2}+{1\over
4g^2}[\Phi,\Phi]^2+{g^2\over 4}F^2 \right\}.
\ee
Also, all of the linear couplings obtained by Schiappa \cite{sch} for the weak
field case, are derivable from the action of equation (\ref{eq:4})
and (\ref{eq:5}). It has been checked that the D1-branes linear
couplings found by Taylor et. al. are included in the non-abelian
action of Myers and the t-duality and s-duality relations are the
same as the ones used by Schiappa. Nevertheless, we have to keep
in mind that there are new couplings not considered before.

Once the relevant action is obtained, we can search for non-commutative
classical solutions. Given the similarity of the mathematical structure
with D-brane physics, we expect to find relevant physical situations
where these types of solutions appear. Nevertheless, in this framework
the building blocks that make the higher dimensional objects are the
``string bits'' of the DLC. Remember that, the matix-value scalar in the
action represent a large number of ``long strings'' and these are the basic
objects that form the higher dimensional branes. Some examples of
non-commutative configurations of strings can be found in
\cite{sch,sil,ben,bert,alf}.

\begin{theacknowledgments}

The author would like to thank the Perimeter Institute for making possible
the conference {\it MRST 2002}, also Alfonso Ramallo, Cesar Gomes, Yoland Lozano,
Joel Rozowsky, Simeon Hellerman and Don Marolf for useful discussions.
This work was supported in part by NSF grant PHY-0098747 to Syracuse
University and by funds from Syracuse University.

\end{theacknowledgments}

\end{document}